\documentclass[12pt]{article}
\usepackage{amsmath,amsthm,amssymb}
\usepackage{graphicx}
\usepackage{psfrag}
\usepackage[letterpaper,margin=1in]{geometry}

\newcommand{\be}{\begin{equation}}
\newcommand{\ee}{\end{equation}}
\newcommand{\bea}{\begin{eqnarray}}
\newcommand{\eea}{\end{eqnarray}}
\newcommand{\beann}{\begin{eqnarray*}}
\newcommand{\eeann}{\end{eqnarray*}}
\newcommand{\benn}{\begin{equation*}}
\newcommand{\eenn}{\end{equation*}}

\def\ra{\rightarrow}
\def\I{\infty}
\def\I{\infty}

\begin{document}

\author{Christian Kuehn\thanks{Institute for Analysis and Scientific Computing, 
Vienna University of Technology, 1040 Vienna, Austria}}

\title{The Curse of Instability}

\maketitle
\textbf{Commentary/editorial to appear in 'Complexity'}

\begin{abstract}
High-dimensional computational challenges are frequently explained via the curse 
of dimensionality, i.e., increasing the number of dimensions leads to exponentially 
growing computational complexity. In this commentary, we argue that thinking on a 
different level helps to understand, why we face the curse of dimensionality.
We introduce as a guiding principle the \emph{curse of instability}, which triggers the 
classical curse of dimensionality. Furthermore, we claim that the curse of instability 
is a strong indicator for analytical difficulties and multiscale complexity. Finally,
we suggest some practical conclusions for the analysis of mathematical models and formulate
several conjectures.  
\end{abstract}

The classical curse of dimensionality \cite{Bellman1} has become synonymous with the difficulty 
of analyzing large data and large mathematical models \cite{Donoho}. The main obstacle is that very few 
additional modelling variables may drastically increase the effective computational cost. 
A typical example is the numerical simulation of coupled interacting nonlinear dynamical systems. Even though a complete 
description of the (probability) density evolution of the system exists via a Liouville 
or Fokker-Planck equation \cite{Gardiner}, these high-dimensional partial differential equations are
often impossible to solve, analytically or numerically. Another example occurs when trying to fit 
models to data, where increasing the dimension of the model, usually makes the input data very 
sparse \cite{Friedman3,VerleysenFrancois}. However, many complex systems seem to defy the curse of 
dimensionality. Although the mathematical model has many dimensions, its effective dimension is significantly 
lower \cite{Antoulas,NadlerLafonCoifmanKevrekidis}. For example, many large-scale chemical reaction networks, such 
as combustion processes \cite{MaasPope}, can frequently be reduced to a few effective slowly-changing dimensions after 
all fast reactions have occurred \cite{ZagarisKaperKaper}. Many formally infinite-dimensional partial differential 
equations are effectively finite-dimensional \cite{Robinson1,Temam}. On the contrary, many simple-looking 
low-dimensional situations exhibit bewildering complexity in the form of chaotic dynamics \cite{Lorenz,Strogatz}. A typical 
physical example is the fluid flow in a long straight pipe, which transitions to turbulence upon increasing the 
fluid velocity \cite{EckhardtSchneiderHofWesterweel}. 

The curse of dimensionality is still viewed in many contexts 
as an open problem. Many attempts have been made to ameliorate its consequences, or even aim to remove it completely, 
in certain contexts \cite{IndykMotwani,KuoSloan,OseledetsTyrtyshnikov,Rust}. However, there is no
solution, which works for all situations. We are really facing a rigid scientific 
barrier in many cases, while it is only a formal or virtual barrier in others.\medskip

Here it is proposed that thinking on a different level is required to understand,
what really triggers or removes this barrier. Just counting 'dimensions' or 'variables'
is insufficient. The true intrinsic curse we are facing is the
\emph{curse of instability}. In fact, we argue below that instabilities (a) cause
an increase in dimensionality, (b) substantially raise the analytical difficulty, and (c) 
are a strong indicator for multiscale dynamical complexity. Of course, it turns out 
that (a)-(c) are intimately related. Although we shall primarily 
illustrate the concepts with examples arising in mathematics and closely-related disciplines, 
it will be shown that the abstract concept occurs, independently, across disciplines. In fact, 
we shall see that the curse of instability has already implicitly triggered the emergence of entirely new 
scientific disciplines. Furthermore, it may lead to formulate more concrete guiding principles 
to address the complexity challenges of the 21st century.\medskip

\textit{(a) Instability causes high-dimensionality:} A paradigm of mathematical modelling for
time-dependent evolving systems is to study the evolution of an initial state $u_0$ of a system at time 
$t=0$ under a mapping $u_{t}=\Phi_t(u_0)$.
Typical examples in this framework are differential equations, iterated mappings, adaptive networks, or in fact
any other dynamical system \cite{GH,Strogatz}. Ideally, we would like $u_t$ to represent the state in a space,
which is tractable for analytical, numerical, and modelling purposes. For example, the population 
density in a city \cite{BrauerCastillo-Chavez} or the voltage inside a neuron 
\cite{ErmentroutTerman} are examples of scalar quantities, 
which are natural choices as variables for certain applications. However, the population density and the 
voltage inside a neuron do not evolve according to a simple universal law given by $\Phi_t$, they
are influenced by more complex multiscale systems they are embedded in, i.e., the global
environmental system \cite{PatzCampbellLendrumHollowayFoley} and the entire brain \cite{SpornsChialvoKaiserHilgetag} 
respectively. Instead of taking into account the 
entire hierarchy of possible models for complex systems, one gains tremendous insight by restricting
or truncating the influences of different levels. For example, in an epidemic in a city, we may just
try to track susceptible or infected population densities \cite{Hethcote}. In the context of a neuron one may just consider
so-called channel gating variables to model ion channels \cite{HodgkinHuxley4}. In this philosophy, we 
neglect many other effects, which are in a suitable sense ``small'', e.g., the smaller-scale network properties for 
epidemics \cite{KeelingEames} or channel noise for neurons \cite{GerstnerKistler}. It turns out that this approach is 
(unreasonably! \cite{Wigner}) effective because small perturbations to the problem formulation are frequently not 
visible in the final result. 

Instability implies that the previous logic is flawed. Indeed, if a system formulation or system 
state changes significantly upon small perturbations, then we have an instability. More formally,
one can define suitable notions of Lyapunov or asymptotic stability for the mathematical model 
\cite{HirschSmaleDevaney}. For example, for an asymptotically globally stable steady state $u^*$ 
we have $\Phi_t(u_0)\ra u^*$ as $t\ra +\I$ for all initial
conditions $u_0$. Classical dynamical systems theory states it may generically happen that such
a situation can be destroyed upon parameter variation \cite{Kuznetsov}. Hence, parameters need to be included as
slow variables in the model \cite{KuehnBook}. In addition, previously neglected effects in the 
\emph{modelling} may suddenly matter near an instability. For example, small random imports of a pathogen 
into a city \cite{ColizzaBarratBarthelemyVespignani} or heterogeneity along a nerve axon \cite{Keener}, 
can suddenly matter. Maybe spatial diffusion of the population inside the city \cite{Mollison} or random 
spikes from other neurons \cite{PlesserGerstner} are also relevant near 
instability? We simply cannot exclude \emph{a-priori} any effects near instability.
The rules what is ``small'' have changed. Adding parameters as new variables, including spatial dimensions, 
as well as introducing stochasticity, just scratch the surface. What about heterogeneity? Non-locality?
Coupling in networks? Microscopic scales? Time-delays? Memory of the system? And the list could be continued
with many other effects. 

This viewpoint shows that instability is a key mechanism: a simple evolution rule fails and 
induces the need for further parametrization and/or re-evaluation of the model complexity. The outcome
is high-dimensionality, i.e., the curse of instability \emph{triggers} the curse of 
dimensionality.\medskip

\textit{(b) Instability induces analytical difficulty:} Away from instabilities, the variety of 
available analytical and numerical techniques is very large. The general philosophy is to 
project, reduce, truncate, coarse-grain, approximate, or otherwise simplify the underlying modelling
as well as the model equations until pen-and-paper mathematical analysis or numerical algorithms 
are feasible \cite{Antoulas,FoiasSellTemam,Kevrekidisetal,Levermore}. However, recent results in 
many scientific disciplines strongly suggest that instability is a key road-block to either apply 
a reduction philosophy or to make it hard to carry out in 
practice \cite{LengauerKinzlerVogelstein,Schefferetal,WilsonSnyderHuysmansMiller}.

To illustrate the mathematical difficulty we consider two recent examples. It is
frequently necessary to model space-time dependent dynamical processes under the influence 
of external stochastic perturbations or to use noise to model fluctuations from microscopic scales.
Although the space-time noise process is usually an infinite-dimensional mathematical object,
there are many situations where it can efficiently be truncated \cite{DaPratoZabczyk}, i.e., certain 
higher-mode components do not play a significant role. However, if instability comes into play, 
then the noise process can be truly infinite-dimensional and tends to effect all modes.
The dramatic increase in mathematical difficulty is exemplified by the fact that only very recently \cite{Hairer1,Hairer2}
a precise mathematical meaning was attached to large classes of stochastic 
partial differential equations\footnote{Martin Hairer received the Fields Medal, (one of) the highest honors
in mathematics, in 2014 for his work on stochastic partial differential equations and the KPZ equation.}. 
A typical example in this class of equations is the Kardar-Parisi-Zhang (KPZ) \cite{KardarParisiZhang} equation, which 
can be used as a model for certain random growth processes, such as burning fronts or crack formation, where the 
interface of the growth process is precisely where the (small-scale) instability occurs \cite{Spohn1}. As a second example, one may consider the notion of hyperbolicity in dynamical systems. Roughly speaking, if a nonlinear system is 
hyperbolic locally, then it is locally possible to predict its behavior from a linearized version of the 
system. The overall system dynamics is stable under small nonlinear perturbations. However, if hyperbolicity 
fails then one is confronted with instability of the mathematical model class. Even many low-dimensional
classes of non-hyperbolic dynamical systems \cite{AvilaLyubichdeMelo,BonattiDiazViana} lead to formidable 
challenges\footnote{Artur Avila received the Fields Medal, (one of) the highest honors in mathematics, 
in 2014 for his work on renormalization in classes of dynamical systems, which are not hyperbolic.}.

A huge variety of further cases could be given to illustrate the difficulty of instabilities.
This includes theoretical as well as experimental issues, side-by-side. Some typical examples are: chaotic dynamics
and turbulence in fluids \cite{HofWesterweelSchneiderEckhardt,PomeauManneville}, microscopic instabilities and 
crack formation in materials \cite{OrtizPandolfi,AbrahamBrodbeckRafeyRudge}, early-warning signs and drastic 
changes in time series \cite{Schefferetal,KuehnCT2}, pattern-formation and Bose-Einstein 
condensation \cite{PitaevskiiStringari,KevrekidisFrantzeskakisCarretero-Gonzalez}, feedback control models 
of the cell cycle \cite{Murray,SerranoHannonBeach}, heterogeneous large-scale networks and financial 
crises \cite{GaiKapadia,MayLevinSugihara}. The list could be continued to staggering length, and effectively 
across all scientific areas \cite{AshwinTimme1}. Whenever one looks at the underlying mathematical models 
formidable challenges occur precisely when instabilities are relevant for the dynamics.\medskip

\textit{(c) Instability indicates multiscale complexity:} Another important issue directly related 
to instabilities are hidden scales. Far away from instabilities, it is 
acceptable to neglect many effects as ``small'' because they occur on scales not relevant for the 
problem at hand. For example, to develop a method to forecast the temperature from today for tomorrow, it
is perfectly safe to ignore the daily change of sea-ice. However, on longer time scales,
the ice-albedo feedback loop \cite{LianCess} can lead to very drastic climate, and hence temperature, changes. 
As a spatial example, one might consider cancer. If a tumor is large, its growth can be described quite 
well by a macroscopic mean-field model on the macroscopic scale of the body \cite{FerreiraMartinsVilela}, while the 
instability and the processes to generate the mutations can occur on significantly 
smaller cellular or even sub-cellular microscales \cite{HanahanWeinberg}.    

The key is that near an instability, the rules of modelling also have changed. Each part of a process 
could contribute significantly. This frequently necessitates to include not only additional small parameters 
but entirely new scales for the problem. Many novel scientific disciplines have emerged 
implicitly due to the necessity for multiscale models. For example, network science \cite{Newman,AlbertBarabasi} 
argues that not only the local individual but also the global interconnected collective scale matters. 
In fact, instability is again at the root of this phenomenon, e.g., a small change in connectivity may 
lead to a large change in global path-length connectivity \cite{WattsStrogatz,Kesten}. The dynamical process 
of adding connections is hence precisely near an instability when the change from local modelling to networks 
arises. A similar effect occurs for coherence or synchronization phenomena on networks, e.g., truly emergent
network effects often arise only upon sufficiently strong coupling \cite{Strogatz2} when each individual node
leaves its intrinsic stable steady state. An analogous argument can be made to explain the emergence of 
systems biology \cite{Kitano}, i.e., the need for a systemic view of biological processes only arises
because of observed instabilities within certain subprocesses.\medskip 

So far, we have demonstrated why the curse of instability is a key factor which triggers high-dimensionality,
analytical difficulty and the need for multiple scales. Clearly, all three effects are related and it is
helpful to use the curse of instability to explain and relate them. The question remains, whether we can
draw any practical consequences from this abstract viewpoint? \medskip 

The first step is to \textit{identify
instabilities}, i.e., to ask whether the process to be modeled does exhibit at the relevant 
observational scale significant qualitative or quantitative changes. From a statistical physics
viewpoint, this will essentially always be the case for ``systems out-of-equilibrium'' \cite{CrossHohenberg}. 
In most complex systems one is usually interested to precisely understand dynamical changes. This is 
a core reason, why practical applications tend to lead to mathematical models, which - 
according to (a)-(c) above - are difficult high-dimensional multiscale problems. 

The second conclusion is the strong necessity to \textit{dynamically simplify models}. Indeed, from (a)-(c) it 
follows that building a global model, which incorporates all possible effects for all parameters and all 
couplings is impossible. Hence, models need to be designed 
and analyzed for specific phenomena and then coupled on hierarchical scales. The influence of smaller
and larger scales have to be simplified on the current scale. For example, in climate models one often introduces 
external forcing \cite{Nicolis} or stochastic fluctuations \cite{NicolisNicolis} to model different scales. 
Of course, this strategy is practiced implicitly by many scientists every day but sometimes 
incorrectly. The point is not to focus purely on simplifying the model to alleviate the computational and analytical 
burden but one should check what dynamical features near instabilities are preserved or lost. 

The third important conclusion
is to improve \textit{bridges between specialist tools}. Near an instability, several effects can be of
equal size, e.g., coupling, noise, and energy dissipation. Of course, one could try to address the mathematical
analysis of this situation just with one particular tool, adapted to cover two additional effects. 
However, this is likely not to alleviate the difficulties. For example, a purely variational approach build upon energy-like
functionals \cite{Reddy} may not easily generalize to include stochasticity \cite{Gardiner} or coupling via complex networks 
\cite{Newman}. Similarly, pure combinatorial network tools or stochastic dynamics techniques may miss important macroscopic 
energy effects. In fact, recent progress in all areas of mathematical theory of complex systems points 
into the direction of better understanding relations between approaches.\medskip

In conclusion, it has been suggested to use the \emph{curse of instability} as another important 
general scientific principle. It has been argued that it triggers the classical curse of dimensionality.
Furthermore, it is a strong indicator for analytical difficulties and multiscale complexity. Although only 
a few practical conclusions based on this principle have been suggested, it is quite reasonable to
assume that far-reaching additional ideas can be developed from the framework sketched here. For example,
based on the curse of instability, one may make the following, quite daunting, conjectures: 

\begin{itemize}
 \item[(C1)] So-called ``universal'' \cite{Kadanoff} or ``generic'' \cite{Wiggins} dynamical principles 
 are always based upon the absence or presence of certain instability classes.
 \item[(C2)] When analyzing big data \cite{Marx} one has to separate the data into regimes with and 
 without instabilities in the underlying process.
 \item[(C3)] Self-organized criticality \cite{BakTangWiesenfeld} and the occurrence of power 
 laws are related because exponential laws usually fail near instability.
 \item[(C4)] Control of complex systems \cite{Siljak} can be ``localized'', i.e., using a different controller near each 
 instability.
 \item[(C5)] The principle of renormalization (or the renormalization group) \cite{Wilson1} can be applied
 successfully because it preserves the underlying stability or instability model class. 
 \item[(C6)] Multiscale computational algorithms \cite{LuskinOrtner1} are only necessary when instabilities occur.
\end{itemize}

Of course, the conjectures (C1)-(C6) are highly speculative. However, they illustrate why
a new guiding principle such as the \textit{curse of instability} can help to pinpoint simple and natural guesses 
to complex scientific problems.\medskip

\textbf{Acknowledgements:} I~would like to thank the Austrian Academy of Sciences ({\"{O}AW}) for 
support via an APART fellowship as well as the European Commission (EC/REA) for support by a Marie-Curie 
International Re-integration Grant.

\end{document}